# Transient Information Adaptation of Artificial Intelligence: Towards Sustainable Data Processes in Complex Projects

Nicholas Dacre [a], Fredrik Kockum [b], & PK Senyo [c]

[a] University of Southampton Business School, Southampton, UK  nicholas.dacre@southampton.ac.uk
[b] University of Southampton Business School, Southampton, UK  f.h.e.Kockum@soton.ac.uk
[c] University of Southampton Business School, Southampton, UK  p.k.Senyo@soton.ac.uk

## Abstract

Large scale projects increasingly operate in complicated settings whilst drawing on an array of complex data-points, which require precise analysis for accurate control and interventions to mitigate possible project failure. Coupled with a growing tendency to rely on new information systems and processes in change projects, 90% of megaprojects globally fail to achieve their planned objectives. Renewed interest in the concept of Artificial Intelligence (AI) against a backdrop of disruptive technological innovations, seeks to enhance project managers' cognitive capacity through the project lifecycle and enhance project excellence. However, despite growing interest there remains limited empirical insights on project managers' ability to leverage AI for cognitive load enhancement in complex settings. As such this research adopts an exploratory sequential linear mixed methods approach to address unresolved empirical issues on transient adaptations of AI in complex projects, and the impact on cognitive load enhancement. Initial thematic findings from semi-structured interviews with domain experts, suggest that in order to leverage AI technologies and processes for sustainable cognitive load enhancement with complex data over time, project managers require improved knowledge and access to relevant technologies that mediate data processes in complex projects, but equally reflect application across different project phases. These initial findings support further hypothesis testing through a larger quantitative study incorporating structural equation modelling to examine the relationship between artificial intelligence and project managers' cognitive load with project data in complex contexts.

**Keywords**: Transient Information, Adaptation, Artificial Intelligence, AI, Sustainable, Data Processes, Complex Projects, Project Management, Project Failure, Project Success, Cognitive Load, Mega Project, Intelligent Systems, Cockpit Design, Megaproject.

## A Backdrop of Dynamic Complex Projectification

In 2007 Crossrail secured a government project for the construction of a new railway line spanning London (Day, 2017). The project was issued a completion date of 2018 at a total cost of £14.8bn. However, due to project deficiencies largely emanating from systems and information failures, project completion has been delayed until 2022 with a reviewed upwards budget of £18.25bn (Cheetham, Moffatt, Addison, & Wiseman, 2019; Sweet, 2019). A 23% increase in spending. The UK's Universal Credit project (Bush, Templer, & Allen, 2019), a government-led consolidation project of disparate benefit payments supported by legacy technologies and information systems (Omar, Weerakkody, & Sivarajah, 2017), was announced in 2010 with a completion date of 2017 at a cost of £2bn.

The new system was initiated in order to modernise and streamline a complex system of benefits, however complex information and software problems have delayed project completion until 2024 with a spiralling budget exceeding £12bn. More recently, the UK HS2 infrastructure project has been announced, linking London with Birmingham, Manchester, and Leeds via high-speed rail with a project budget of £56bn and an initial completion date of 2026 (Charlson & Dunwoody, 2018). It is now anticipated to complete in 2040 at a cost of £106bn due in part to a lack of understanding and comprehension of the complexity of the project. These examples follow a global trend, where megaprojects have an estimated failure rate as high as 90% (Lenfle & Loch, 2017). However, small and medium projects are not immune to project failure with an estimated 80% of projects failing to wholly meet their planned objectives (APM, 2015).

Despite these indicative high failure rate in projects, the increasing projectification of services, products, innovations, and processes in organisations and society is a major contributor to the UK economy. The practice contributes around £156.5bn of annual GVA and provides employment for an estimated £2.13m full-time equivalent workers (APM, 2019). Therefore, when projects fail, this can have catastrophic effects on the economy, society, and the environment.

The concept of Artificial Intelligence (AI) as coined by John McCarthy during the Dartmouth Conference in 1956 (Andresen, 2002), reflects an increasing convergence of technology, innovation, and human reasoning. Although AI has been conceptualised, adapted, and re-appropriated, it is through the onset of recent incremental yet salient technological and systems developments that the field is permeating into a range of sectors and industries. For example, AI is making further in-roads into Sports (Chu, Shih, Chou, Ahamed, & Hsiung, 2019), Construction (Juszczyk, 2017), Entertainment (Meena, Jingar, & Gupta, 2020), Healthcare (Jiang et al., 2017), and Project Management (Dacre, Senyo, & Reynolds, 2019; Wauters & Vanhoucke, 2016). The project management field is showing promise in cognitive adaptations, and transient information, where access to data, retention, and execution is important across different stages of the project lifecycle.

Project data attributes can span a number of different salient dimension, such as complex, structured, unstructured, volume, velocity, and variety (Whyte, Stasis, & Lindkvist, 2016). Project managers operating in increasing challenging contexts, require further insights to manage complex project data (Pedroza–SN, 2019). Therefore, leveraging nascent disruptive technology attributes such as AI which may on one hand aid project management cognitive processes, however the rate of development is increasingly proving challenging for project practitioners (Bhavsar, Shah, & Gopalan, 2019; Meharwade et al., 2019). Elements of technological advancements, and innovation are spreading at a faster rate than practitioners are able to fully benefit from their implementation. Furthermore, organisations are aware of a number of nascent innovations (i.e. innovations which are gaining popularity but yet to be fully developed for project purposes), however have limited capacity and ability to adapt to these dynamic technologies.

Cognitive load, defined as one's ability to process and apply knowledge (Paas, Tuovinen, Tabbers, & Van Gerven, 2003), or knowledge-based attributes may span tacit and explicit issues which include knowledge management, transfer, coordination, sharing, access, and intelligence attrition throughout the project lifecycle. These play a part in the overall process of projects and portfolio management (Lin, Müller, Zhu, & Liu, 2019; Wei & Miraglia, 2017). Hence the need for project research to investigate the role of AI with complex data, to offer a model for project organisations to best leverage these for project excellence (Sirisomboonsuk, Gu, Cao, & Burns, 2018).

## Research Outline

This research in development employs an exploratory sequential mixed method. This approach supports the analysis of emergent themes from qualitative data. These themes are then used to guide the development of a large-scale quantitative survey for analysis and discussion (Creswell, 2013; Denzin & Lincoln, 2005; Grbich, 2012). Mixed methods are particularly adept in uncovering emergent themes, exploring complex ideas, and understanding complex issues (Creswell, 2013; Reynolds & Dacre, 2019). An initial round of qualitative empirical data has been collected through a number of semi-structured interviews with domain experts. This has helped guide early thematic interpretations of transient information adaptions of artificial intelligence in complex projects.

Semi-structured interviews were recorded with individual consent and adhered to ethical guidelines. Respondents were able to terminate interviews at any point. Interviews were transcribed as close as possible to the interviews in order to capture the depth of the insights and were cross-checked against any additional contextual notes acquired during the interview. Coding of the interviews followed a structured iterative six phase approach, which includes data familiarisation, initial coding, searching for themes, defining and naming themes, and producing a summary data report (Braun & Clarke, 2006).

## Brief Discussion

Megaprojects are indicative of extreme levels of complexity (Davies, 2019). Brookes et al. (2017) state the nature of the long-term duration of mega-projects offers a novel way of researching the multiple temporalities within long-term projects. Project managers in these environments have a greater degree of confidence for failure than success. Furthermore, historical examples of successful megaproject, such as the Manhattan and Apollo project remain few and far between (Flyvbjerg, 2017). Hence project managers have a larger data set of dependant failure rates than success factors. As such, modern mega projects which rely on increasingly complex data-points for design, definition, planning, implementation and control, require comparative levels of cognitive processing and accuracy.

Humans are generally very good at managing cognitive load efficiently (Paas et al., 2003; Sweller, 2011). In part this is what has led to the development of modern society. Our ability as a species to acquire knowledge, construct critical thoughts, and apply our knowledge and skills in constructive ways. However, increasing cognitive loads can lead to process paralysis (Kirsh, 2000). For example a trained pilot is able to cogently recognise a number of key patterns in ensuring flight conditions are optimal and sustained across an array of data-points (Tudoreanu, 2016). However, in sever conditions such as an emergency for which the pilots have no prior training, this leads to a cognitive overload, where the individual's ability to recognise, assimilate, and process information in a cogent way, becomes diminished as the velocity and variety of data increases (Puma, Matton, Paubel, Raufaste, & El-Yagoubi, 2018). This is why cockpit design is fundamental to ensuring the safety of the plane. By focusing on cognitive load and transient information processes mediated through intelligent systems, aeroplane manufacturers enhance the ability and capacity of pilots to remain cogent with fast-paced changing dynamics. Project managers operating in complex megaprojects, may face a similar cognitive overload which can diminish their ability to make cogent decisions, leading to process paralysis. The concept of AI in this research study may thus be framed as a cognitive load enhancer through transient information adaptations in complex projects. Therefore the aim of the study is to further elaborate and test this overarching hypothesis.

## Work in Progress

A second round of qualitative based interviews to further refine early insights will take place with 25 domain experts or project professionals. These will be targeted based on experience, qualifications, and their role throughout the project lifecycle. These second round of interviews will adopt the initial qualitative approach, in that to extract rich meaning from the data in context, interviews will be scheduled preferably in person, however where this is not feasible using digital facilitated communication platforms. Each interview will be broadly themed and guided by the overarching research enquiry, with a duration between 1 and 2 hours.

The third phase of the empirical data collection will involve the design and implementation of an online survey guided by the emergent themes from the qualitative approach. The survey will focus on examining mediating variables. The survey will be distributed to 400 domain experts. Survey questions will initially establish key indicators, such as rank, role, responsibilities, experiences, specific area of expertise, projects managed, industry, salary range, number of projects managed, failed projects, reasons, successful projects, and constraints. This is not an exhaustive list but illustrates key indicators. The survey will be managed through the use of the Qualtrics survey software. Quantitative data collection will follow established academic ethical guidelines at all times to ensure informed consent and uphold the safety and anonymity of the interviewees, respondents, and individuals undertaking the research process. The results of the empirical survey will then be fully analysed using Structural Equation Modelling (SEM).